\documentclass[acmsmall]{acmart}

\setcopyright{rightsretained}
\acmDOI{10.1145/3643742}
\acmYear{2024}
\copyrightyear{2024}
\acmSubmissionID{fse24main-p240-p}
\acmJournal{PACMSE}
\acmVolume{1}
\acmNumber{FSE}
\acmArticle{16}
\acmMonth{7}
\received{2023-09-29}
\received[accepted]{2024-01-23}

\newcommand{\code}[1]{{\small \texttt{#1}}}

\usepackage{graphicx}
\usepackage{tikz}
 
\usepackage{multirow}
\usepackage{listings}
\usepackage{tcolorbox}
\usepackage{color}
\usepackage{xcolor}
\usepackage{graphicx}

\usepackage{algorithmic}
\usepackage[ruled,linesnumbered,vlined]{algorithm2e}
\usepackage{setspace}
\usepackage{listings}
\usepackage{multicol}
\usepackage{makecell}
\usepackage{amsthm}
\usepackage{enumitem}
\usepackage{tikz}
\usepackage{xspace}
\usepackage{wrapfig}
\usepackage{calc, scalerel}
\usepackage{mathtools}
\usepackage{caption}
\usepackage{subcaption}
\usepackage{colortbl}
\usepackage{bbm}
\usepackage{pifont}
\usepackage{hyperref}

\definecolor{celadon}{rgb}{0.67, 0.88, 0.69}
\definecolor{codegreen}{rgb}{0,0.6,0}
\definecolor{codegray}{rgb}{0.5,0.5,0.5}
\definecolor{codepurple}{rgb}{0.58,0,0.82}

\lstdefinestyle{mystyle}{
    commentstyle=\color{codegreen},
    keywordstyle=\color{magenta},
    numberstyle=\tiny\color{codegray},
    stringstyle=\color{codepurple},
    basicstyle=\ttfamily\footnotesize,
    breakatwhitespace=false,
    breaklines=true,
    captionpos=b,
    keepspaces=true,
    numbers=none,
    numbersep=5pt,
    showspaces=false,
    showstringspaces=false,
    showtabs=false,
    tabsize=2
}
\newcounter{findingCounter}
\newcommand{\name}{DyPyBench}
\newcommand{\dynamic}{$G_{dynamic}$}
\newcommand{\static}{$G_{static}$}

\begin{document}

\title{DyPyBench: A Benchmark of Executable Python Software}

\author{Islem Bouzenia}
\orcid{0000-0002-3920-3839}
\affiliation{%
	\institution{University of Stuttgart}
	\city{Suttgart}
	\country{Germany}
}
\email{fi\_bouzenia@esi.dz}

\author{Bajaj Piyush Krishan}
\orcid{0009-0003-1227-2265}
\affiliation{%
	\institution{University of Stuttgart}
	\city{Stuttgart}
	\country{Germany}
}
\email{st173644@stud.uni-stuttgart.de}

\author{Michael Pradel}
\orcid{0000-0003-1623-498X}
\affiliation{%
	\institution{University of Stuttgart}
	\city{Suttgart}
	\country{Germany}
}
\email{michael@binaervarianz.de}


\begin{abstract}
Python has emerged as one of the most popular programming languages, extensively utilized in domains such as machine learning, data analysis, and web applications.
Python's dynamic nature and extensive usage make it an attractive candidate for dynamic program analysis.
However, unlike for other popular languages, there currently is no comprehensive benchmark suite of executable Python projects, which hinders the development of dynamic analyses.
This work addresses this gap by presenting \name{}, the first benchmark of Python projects that is large-scale, diverse, ready-to-run (i.e., with fully configured and prepared test suites), and ready-to-analyze (by integrating with the DynaPyt dynamic analysis framework).
The benchmark encompasses 50 popular open-source projects from various application domains, with a total of 681k lines of Python code, and 30k test cases.
\name{} enables various applications in testing and dynamic analysis, of which we explore three in this work:
(i) Gathering dynamic call graphs and empirically comparing them to statically computed call graphs, which exposes and quantifies limitations of existing call graph construction techniques for Python.
(ii) Using \name{} to build a training data set for LExecutor, a neural model that learns to predict values that otherwise would be missing at runtime.
(iii) Using dynamically gathered execution traces to mine API usage specifications, which establishes a baseline for future work on specification mining for Python.
We envision \name{} to provide a basis for other dynamic analyses and for studying the runtime behavior of Python code.
\end{abstract}


\maketitle


\section{Introduction}

Python has grown rapidly in recent years to become one of the most frequently used programming languages in a variety of fields, including machine learning, data analysis, and web applications.
For example, a 2022 report on programming languages used in projects hosted on GitHub lists Python as the second-most popular of all languages.\footnote{\url{https://octoverse.github.com/2022/top-programming-languages}}
The success of Python can be attributed to its simplicity, flexibility, and extensive collection of libraries and frameworks.
Python's dynamic nature, which allows for dynamic type checking and modification of object structures during runtime, further enhances its appeal to developers.
 
The dynamic nature of Python also presents challenges when it comes to program analysis, error detection, security assessment, and performance optimization.
Various dynamic program analysis techniques have been developed to address these challenges, e.g., to detect bugs~\cite{Xu2016a}, to enforce differential privacy~\cite{DBLP:conf/csfw/AbuahSDN21}, to slice programs~\cite{DBLP:conf/compsac/ChenCZXCX14}, or to understand the performance of a program via profiling~\cite{berger2023triangulating}.
These techniques play a crucial role in identifying programming errors, security vulnerabilities, and performance bottlenecks.
To help develop, evaluate, and compare dynamic analysis techniques, other popular programming languages have benchmark suites of executable code.
For example, DaCapo for Java~\cite{Blackburn2006}, SPEC CPU for C/C++~\cite{Henning2006}, and Da Capo con Scala~\cite{sewe2011capo} have contributed significantly to work by both researchers and practitioners on their respective programming languages.

Despite the popularity and importance of Python, there currently is no equivalent benchmark suite for Python, which hinders the progress and evaluation of dynamic analyses.
Instead, researchers, who want to work on a novel dynamic program analysis for Python, would like to study the runtime behavior of Python code, or want to gather data from executions of Python software, must spend significant efforts on creating a suitable set of executable programs.
As a result, many evaluations that involve dynamic analysis of Python are performed on a relatively small set of benchmark programs, e.g., nine projects~\cite{fse2022-DynaPyt}, five projects~\cite{fse2023-LExecutor}, six projects~\cite{chen2014dynamic}, and eleven projects~\cite{Xu2016a}.
Moreover, since every team working on Python creates their own benchmark, it is difficult to compare different techniques and results with each other.
 
In response to this gap, this paper presents \name{}, a novel benchmark of executable Python projects.
We envision our benchmark to facilitate future work on dynamically analyzing Python.
To this end, the benchmark is the first to offer four important properties -- large-scale, diverse, ready-to-run, and ready-to-analyze -- as described in the following:
\begin{itemize}
	\item \emph{Large-scale}.
	To offer users an extensive collection of executable code, the benchmark includes a substantial number of popular and non-trivial projects.
	The definition of ``large-scale'' may vary depending on the specific use case.
	In this work, we have curated a dataset comprising 50 projects, collectively amounting to 681k lines of code.
	This size is an order of magnitude larger than the benchmarks considered in previous evaluations, as listed above.
	
	\item \emph{Diverse}.
	Given Python's numerous applications, it is critical that a benchmark covers a wide number of application domains, which each have their own code style and libraries.
	\name{} attempts to represent the breadth and diversity of Python usage by including projects systematically sampled from different application domains, enabling extensive assessments and studies of code behavior across different contexts.
	
	\item \emph{Ready-to-run}.
	Executing the code of a real-world, complex project is non-trivial, as it requires a suitable project configuration, installing third-party dependencies, and inputs that exercise the code.
	The effort spent on setting up projects can be significant, especially for users who are not familiar with the project, and usually, this effort needs to be spent again and again for each new project.
	\name{} comes with all projects already set up and uses the test suites of these projects to provide inputs.
	To facilitate running the code irrespective of variations in project dependencies, configurations, and execution environments, \name{} provides a unified interface that allows users to run the benchmark with a single command.
	Moreover, we provide a Docker image that encapsulates the benchmark and all its dependencies, ensuring reproducibility and longevity.
	
	\item \emph{Ready-to-analyze}.
	To facilitate the dynamic analysis of the projects in \name{}, the benchmark integrates with DynaPyt~\cite{fse2022-DynaPyt}, a general-purpose dynamic analysis framework for Python.
	The integration allows for instrumenting and analyzing projects with a single command, making it straightforward to apply a new dynamic analysis to a wide range of projects.
\end{itemize}

To evaluate the usefulness and practicality of \name{}, we perform a series of experiments covering a wide range of research areas.
In particular, we apply our benchmark in three usage scenarios:
\begin{enumerate}
	\item We use \name{} to empirically compare statically and dynamically created call graphs.
	To this end, we use the dynamic analysis infrastructure built into \name{} to gather dynamic call graphs and compare them to the results of a state of the art static call graph generator~\cite{DBLP:conf/icse/SalisSLSM21}.
	Our results help understand the strengths and weaknesses of both approaches, and provide some guidelines for future work on call graph construction.
	\item We use \name{} to create training data for LExecutor~\cite{fse2023-LExecutor}, a recent neural network-based technique for learning-guided execution.
	In this scenario, we apply the existing source code instrumentation tool provided by LExecutor to the code in \name{} and then execute our benchmark.
	By using the resulting 436,355 data examples  as training data for the LExecutor model, we find that more data helps improve the accuracy of the model.
	\item Finally, we use \name{} to mine specifications from execution traces.
	For this application, we again build upon the dynamic analysis support built into our benchmark, this time to extract traces of function calls. \name{} provided thousands of sequences of calls that allowed to extract some meaningful patterns, which sets a baseline for future work on dynamic specification mining for Python.	
\end{enumerate}
For all three of these applications, one would usually have to spend a significant effort on setting up suitable projects and finding inputs, e.g., in the form of test suites, for exercising their code.
Instead, \name{} provides this setup, which hugely facilitates these and future dynamic analyses for Python.
 
In summary our work contributes the following:
\begin{itemize}
\item We address the lack of a comprehensive benchmark suite of executable Python projects by creating \name{}.

\item We integrate our benchmark with the general-purpose dynamic analysis framework DynaPyt~\cite{fse2022-DynaPyt}, which allows for performing arbitrary dynamic analyses on the entire benchmark with minimal effort.

\item We illustrate the usefulness of \name{} in three application scenarios.
\end{itemize}

\section{Methodology}

This section describes our methodology for creating a benchmark of executable Python software.
We begin by presenting the criteria for selecting projects to include in the benchmark, and then describe our methodology for making these projects ready-to-execute and ready-to-analyze.

\subsection{Selecting Projects}
\label{sec:selecting}

To ensure the diversity and representativeness of the benchmark, we select projects from the Awesome Python repository,\footnote{\url{https://github.com/vinta/awesome-python}} which is a curated list of popular, open-source Python projects.
The Awesome Python repository has 173k stars on GitHub and more than 400 contributors, which are indicators of its popularity and adoption by the community.
The curated list contains 679 projects, including libraries, frameworks, and applications, which are classified into 90 categories.

We select a subset of the projects from the curated list, following three criteria designed to ensure the diversity, quality, and representativeness of our benchmark.
First, to increase the quality and relevance of the benchmark, we consider only projects that have at least 500 stars on their respective GitHub repositories.
GitHub stars serve as an indication of a project's popularity, reflecting its usefulness and community support.
Second, to make \name{} diverse, we sample projects from different categories in the Awesome Python list.
Each category covers a different application domain, such as web crawling, machine learning, and robotics.
We randomly sample from all projects in the Awesome Python list, with the constraint to pick at most one project from each category.
Finally, as our goal is to create an executable benchmark, we focus on projects with test suites. Specifically, we focus on test suites that can be executed with pytest, i.e., one of the most popular testing frameworks for Python.
The selected tests include tests written based on Python's built-in \code{unittest} framework, as well as tests written with pytest itself.

Alternative to sampling projects from the Awesome Python list, we could have selected the most downloaded projects from the Python Package Index (PyPI).
However, we observe that the most downloaded projects do not fully represent the diversity of the Python ecosystem, but are biased toward kinds of projects that many others depend on, such as tools to build and set up projects.
Instead, the Awesome Python list provides an independently curated list of projects grouped into a diverse set of application domains.
A downside of using the Awesome Python list is that it focuses on open-source projects only, and hence, may not be fully representative of the entire Python ecosystem.
We decided to focus on open-source projects, as they are more likely to be used in research and are more accessible to the community.

\subsection{Enabling Execution}

As executability is an important property of our benchmark, we invest significant efforts into automating the process of setting up the projects and their test suites.
Our benchmark can be set up in two ways.
First, we provide a Docker image that encapsulates the benchmark and all its dependencies.
This image can be used to run the benchmark on any machine that supports Docker, ensuring reproducibility and longevity.
Second, we provide a command-line interface that allows users to automatically install and set up the benchmark and its projects.
This process involves cloning the projects, installing their dependencies, and configuring their test suites.
To the extent possible, \name{} invokes standard Python tools for installing, building, and testing projects.
However, some projects require various non-standard steps, such as installing unspecified third-party dependencies or slightly adapting the developer-provided test suites.
To address these cases, we manually inspect the projects and make the necessary adjustments to ensure that they are ready-to-execute.
From the perspective of a user of our benchmark, this process is transparent, as the benchmark provides a single command for setting up and executing all or individual projects.

The following describes the steps of our automated setup process in detail.
At first, we clone the project repositories from GitHub.
To ensure reproducibility, we clone the repositories at a specific commit, which is the latest commit on or before January 18, 2023.
Then, we create a Python virtual environment for each project, which ensures that the project's dependencies are installed in an isolated manner.
Next, we install the project's dependencies within the virtual environment.
If available, this step uses the project's \code{requirements.txt} file, which we augment with additional dependencies if necessary.
Finally, we install the project itself, ensuring that it is properly set up within its dedicated virtual environment.

To ensure that the test suites of the projects are ready-to-execute, we configure them to run with pytest.
This step involves specifying the locations of the test cases within the project.
For some projects, additional dependencies may be required to execute the test suite.
These dependencies are often not mentioned in the requirements file but can be found in the project's README instructions.
To handle this scenario, we manually collect the necessary dependencies for each project.
Furthermore, we overwrite specific test files of some projects, e.g., for tests that run into an infinite loop or that consume an unacceptably high amount of memory.
This affects a total of 51 test files across 18 projects.
Finally, we validate that the test suites are ready-to-execute by running all the above steps on a fresh machine and by checking that the tests execute successfully.

\begin{table*}[t]
	\caption{Projects in the benchmark.}
	\label{tab:projects}
	\setlength{\tabcolsep}{4pt}
	\small
	\begin{tabular}{@{}ll|ll|ll@{}}
		\toprule
		Project & Domain & Project & Domain & Project & Domain \\
		\midrule
		akshare & Downloader & grab & Web crawling & python-decouple & Configuration  \\
		arrow & Date \& time & graphene & GraphQL & python-diskcache & Caching \\
		black & Code formatter  & gunicorn & WSGI servers & python-future & Compatibility \\
		blinker & Misc.  & html2text & Web content & python-patterns & Algorithms \\
		supervisor & DevOps & lektor & Static sites & PythonRobotics & Robotics \\
		celery & Task queues & marshmallow & Serialization & pyvips & Parallel img. proc. \\
		cerberus & Data validation & mezzanine & CMS & requests & HTTP clients \\
		click & CLI app dev. & moviepy & Video & schedule & Job schedul. \\
		code2flow & Code analysis & pdoc & Documentation & seaborn & Data visual. \\
		delegator.py & Processes & pickledb & Database & streamparse & Distr. comput. \\
		dh-virtualenv & Distribution & pillow & Image proc. & structlog & Logging \\
		elasticsearch-dsl & Search  & pudb & Debugging & thefuck & Cmd-line tools \\
		errbot & ChatOps tools  & pydub & Audio & uvicorn & ASGI servers \\
		flask-api & RESTful API & pyfilesystem2 & Files & webassets & Web asset mgmt. \\
		funcy & Functl.\ progr. & pyjwt & Authentic. & wtforms & Forms  \\
		furl & URL manip. & pypdf & Format proc.  & zerorpc-python & RPC servers \\
		geopy & Geolocation & pyquery & HTML manip. \\
		\bottomrule
	\end{tabular}
\end{table*}

\subsection{Enabling Dynamic Analysis}
\label{sec:dynamic analysis}

One of the main goals of this work is to facilitate dynamic analysis of Python projects.
To this end, the benchmark integrates with DynaPyt~\cite{fse2022-DynaPyt}, a general-purpose dynamic analysis framework for Python.
DynaPyt supports a range of analyses by providing an API of callbacks that are triggered whenever a specific kind of runtime event occurs, such as a function call, the creation of a new object, or a binary operation.
We show two DynaPyt-based applications of \name{} in Sections~\ref{sec:call graphs} and~\ref{sec:spec mining}.

Performing a dynamic analysis with DynaPyt requires instrumenting the source code of the target project.
\name{} facilitates this step by providing a single command for instrumenting all or selected projects in the benchmark.
The command applies DynaPyt's instrumentation tool to the source code of the selected projects.
\name{} offers two variants of this command:
One variant instruments all Python code in a project, which is useful to analyze the executions of both the tests and the project's main code.
The other variant instruments only the application code, i.e., excluding the test suite.
We use the latter variant, e.g., for performing a dynamic call graph analysis (Section~\ref{sec:call graphs}).
Finally, \name{} provides a command for running the instrumented projects, which automatically invokes the callbacks into a user-provided dynamic analysis whenever a runtime event occurs that is of interest to the analysis.

Beyond dynamic analyses built with DynaPyt, the benchmark can, of course, also be used with other dynamic analysis tools.
For example, one of our applications (Section~\ref{sec:lexecutor}) is based on a custom, instrumentation-based dynamic analysis.
Furthermore, the benchmark could also be easily run with dynamic analysis tools that do not require source-level instrumentation, but that instead build upon Python's built-in tracing library \code{sys.settrace}.


\section{\name{}}

By applying the methodology described in the previous section, we create \name{}, a large-scale, diverse, ready-to-run, and ready-to-analyze benchmark of Python projects.
This section presents the composition of the benchmark (Section~\ref{ss:composition}), its runtime properties (Section~\ref{ss:runtime}), and details of its implementation (Section~\ref{ss:implementation}).

\subsection{Composition of the Benchmark}
\label{ss:composition}

\name{} consists of 50 projects, listed in Table~\ref{tab:projects}.
The code in these projects sums up to 681k lines of Python code, which can be considered large-scale in comparison to current community standards.
For comparison, recent papers on dynamic analyses for Python are usually evaluated with an order of magnitude fewer projects, e.g., nine projects~\cite{fse2022-DynaPyt}, five projects~\cite{fse2023-LExecutor}, six projects~\cite{chen2014dynamic}, and eleven projects~\cite{Xu2016a}.
%
Each project in \name{} covers a different application domain, where the domains are determined by the Awesome Python repository (Section~\ref{sec:selecting}).
Table~\ref{tab:projects} lists the 50 domains along with the projects included in the benchmark.
Within the selected projects, 94.5\% of files are Python files, whereas some files are written in other  languages (2.0\% JavaScript, 1.2\% HTML, 0.8\% C).

\subsection{Runtime Properties of the Benchmark}
\label{ss:runtime}

\begin{table}[t]
	\caption{Properties of \name{}.}
	\label{tab:properties}
	\setlength{\tabcolsep}{20pt}
	\begin{tabular}{@{}lr@{}}
		\toprule
		Metric & Value \\
		\midrule
		Projects & 50 \\
		Lines of code &681k \\
		Test cases: \\
		\hspace{1em} Total & 29,511 \\
		\hspace{1em} Passing &27,569 \\
		\hspace{1em} Failing & 270 \\
		\hspace{1em} Skipped & 1,672 \\
		Lines of executed code: \\
		\hspace{1em} Total lines & 558k \\
		\hspace{1em} Coverage & 82\% \\
		Execution time: \\
		\hspace{1em} Avg./project & 71 seconds \\
		\hspace{1em} Min. & 1 seconds \\
		\hspace{1em} Max. & 1,362 seconds \\
		\hspace{1em} Total & 3,568 seconds \\
		\bottomrule
	\end{tabular}
\end{table}

Because a key property of \name{} is to be executable, the following takes a deeper look into the runtime properties of the benchmark.
Table~\ref{tab:properties} summarizes the key figures.
The projects in \name{} come with a comprehensive set of test cases, totaling 29,511.
The number of test cases per project ranges from 1 to 3,947, with an average of 590 test cases per project.
\begin{figure}
	\centering
	\includegraphics[width=1.01\linewidth]{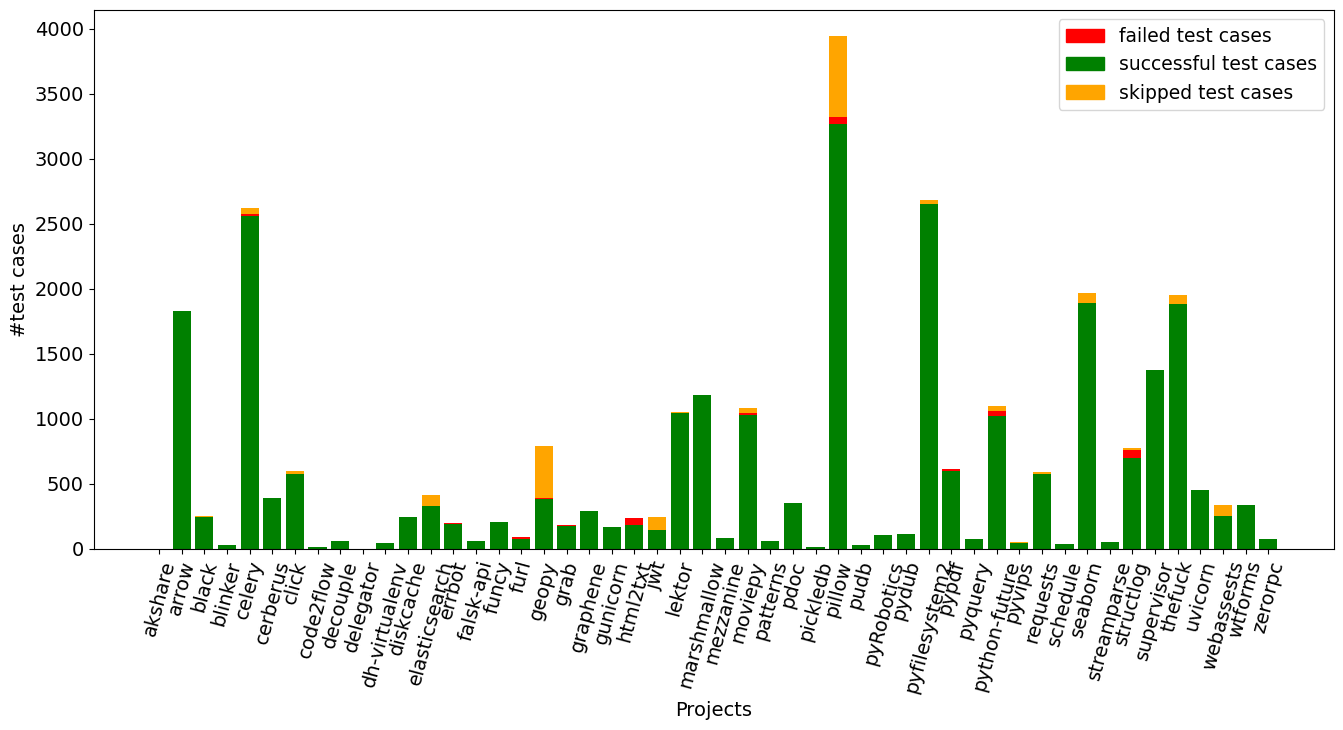}
	\caption[]{Number of successful, failed, and skipped test cases for each project of the benchmark.}

	\label{fig:testcasescover}
\end{figure}

\begin{figure}
	\centering
	\includegraphics[width=0.5\linewidth]{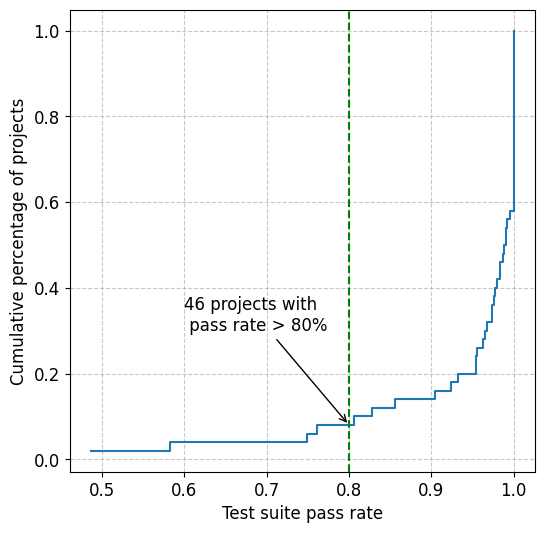}
	\caption{Cumulative distribution of test suite pass rates for all projects.}
	\label{fig:passratecdf}
\end{figure}

Out of all test cases in the benchmark, 27,569 test cases pass, while 270 test cases fail and the rest gets skipped.
\name{} has 31 projects with no failing test cases, while 13 projects have between 1 and 10 failing tests.
Figure~\ref{fig:testcasescover} shows the number of successful, failing and skipped test cases per project.
The pass rate, i.e., the percentage of passing test cases among all test cases, across all projects in \name{} is 93\%.
To better understand the distribution of test case outcomes across projects, Figure~\ref{fig:passratecdf} shows the cumulative distribution of test case pass rate.
The plot shows that 41 projects have a pass rate exceeding 90\%, and 46 projects have 80\% or more test cases passing.
This high pass rate not only allows users of our benchmark to execute a lot of code, but also to observe mostly valid executions.

There are two main reasons why some projects have failing test cases.
First, some tests fail due to incompatible hardware.
For example, some test cases are designed to run on a specific hardware, e.g., a GPU or a specific CPU architecture.
Our Docker-based setup provides software abstraction, but not hardware abstraction, which can lead to test failures depending on the used machine.
Second, some tests fail due to incompatible or unavailable software.
Examples include test cases designed to run on a particular operating system and test cases that use online services that require authentication.
Because our Docker-based setup provides a single consistent software environment, including a specific operating system, tests that require another software environment are doomed to fail.

\begin{figure}
	\centering

	\includegraphics[width=1.01\linewidth]{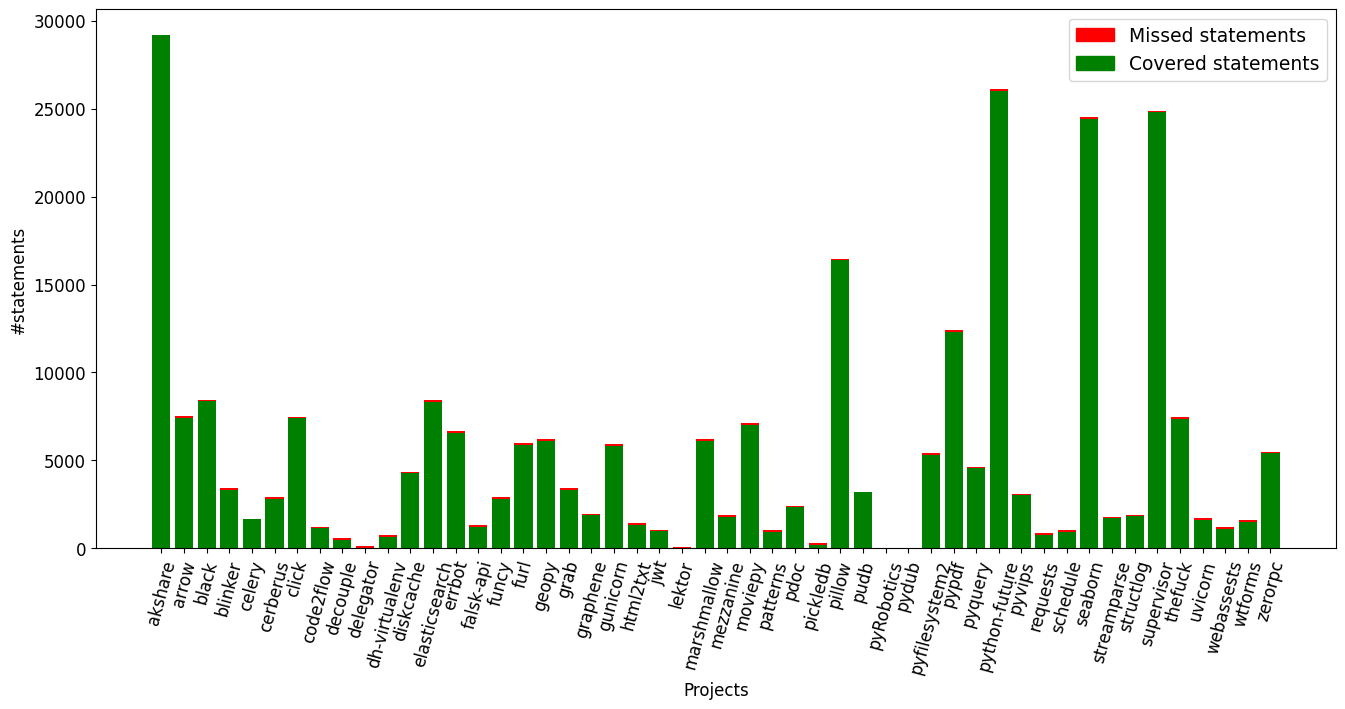}
	\caption{Distribution of covered statements vs uncovered statements during the execution of test suites for each project.}
	\label{fig:coverage}
\end{figure}

Figure~\ref{fig:coverage} shows the distribution of covered and uncovered statements as a result of test execution for each project. The figure illustrates that the number of missed statements is noticeably low for the vast majority of the projects. Furthermore, 27 projects have coverage levels that exceed 85\%. This high coverage not only demonstrates comprehensive testing within our benchmark, but it also plays an important role in providing diverse and meaningful data for a variety of applications, as discussed in more detail in Section~\ref{sec:applications}.

\begin{figure*}[t]
	\centering
	\includegraphics[width=1.01\linewidth]{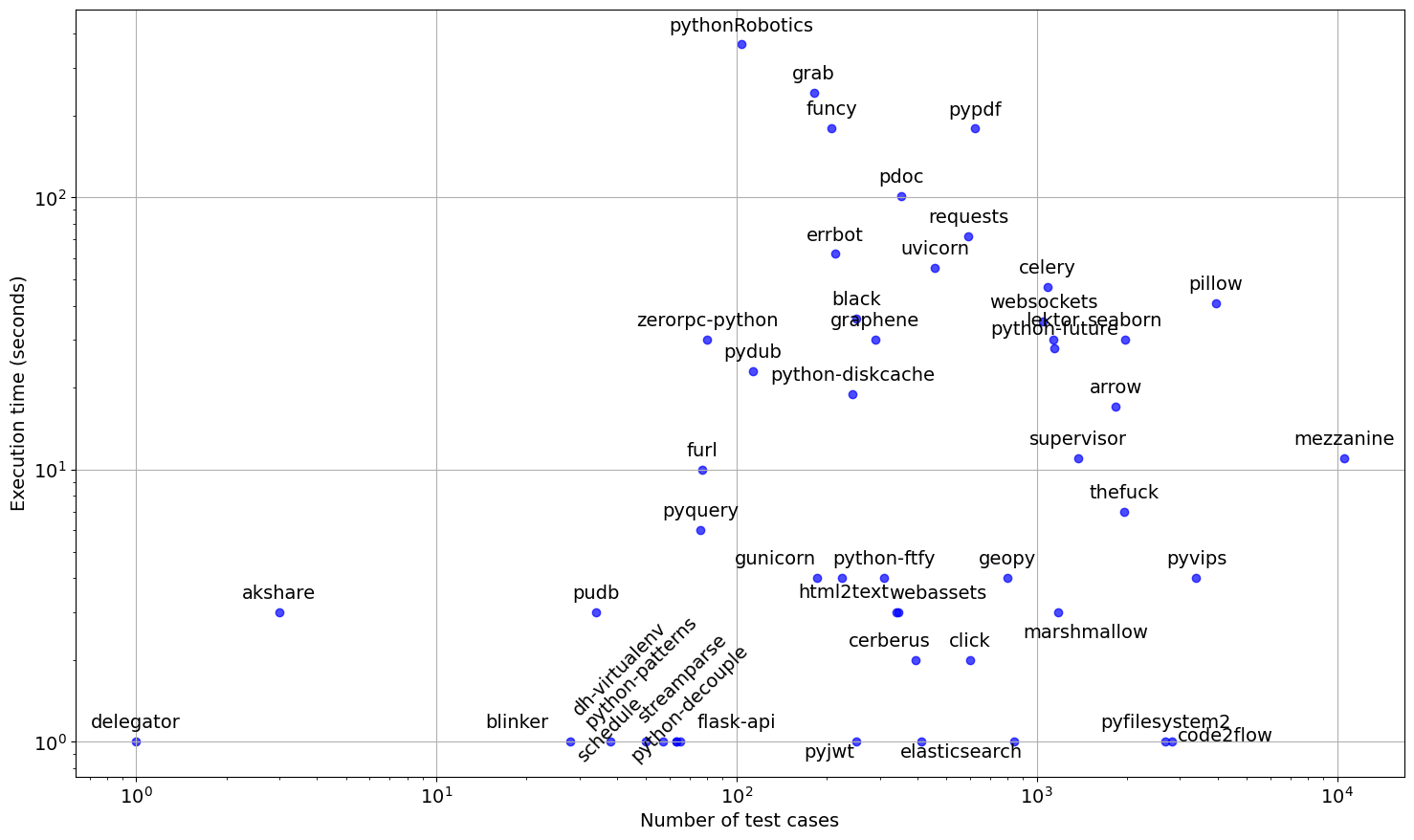}
	\caption{Number of test cases vs. execution time for each projects.}
	\label{fig:projectstesttime}
\end{figure*}

Finally, we report the wall-clock time required to execute the test suites of the projects in \name{}.
The following numbers are obtained on a standard computer with an Intel Xeon CPU running at 2.10GHz and 32GB of memory. We launch all the test suites on a single CPU core.
The time to execute the test suite of single project ranges from 1 second to 1,362 seconds. Figure~\ref{fig:projectstesttime} illustrates the runtime for each project (rounded to full seconds), as well as the corresponding number of test cases.
The total runtime for all projects combined amounts to 3,568 seconds, i.e., about one hour.
The average time required to execute a complete test suite is 71 seconds.
Notably, 39 projects can complete their test suite within 50 seconds or less, while there are four projects with runtimes exceeding 200 seconds.
The diversity in project sizes and runtimes of their corresponding test suites allows users of \name{} to sample projects with different characteristics, depending on the specific needs.

\subsection{Implementation}
\label{ss:implementation}

Inspired by other widely used benchmarks in the community~\cite{Just2014,DBLP:journals/corr/abs-1903-06725}, \name{} comes with a convenient command-line interface that allows users to interact with the benchmark and its projects.
For example, the interface supports running all or selected projects, instrumenting their source code, or applying a dynamic analysis to them.
For reproducibility and to ensure the longevity of the benchmark, we provide an Ubuntu 20.04-based Docker image that encapsulates the benchmark and all its dependencies.
We use Python~3.10 as the default Python version across all our Python environments, considering its widespread usage and stability.
To support dynamic analysis, the benchmark integrates with DynaPyt version 0.3.1.

Currently, \name{} contains 50 projects, but it is designed to be easily extensible.
The benchmark can be extended by adding the GitHub URL of new projects to a specific text file.
Our installation script will try to install and configure the new projects based on frequently observed setup patterns.
In case our script fails to automatically install a project, the user can edit the installation script to add a special case.

All our experiments are conducted on a machine with an Intel Xeon Gold 6230 CPU, 64GB RAM limit and 300GB disk space.
The disk space is needed only when performing the applications described below on all projects at once.

\section{Applications}
\label{sec:applications}

We demonstrate the usability and usefulness of DyPyBench by studying three applications of the benchmark. 
First, we compare dynamic and static call graphs extracted for the programs in DyPyBench (Section~\ref{sec:call graphs}). 
Second, we use the benchmark to generate training data for LExecutor~\cite{fse2023-LExecutor}, a machine learning-based technique for executing incomplete code snippets (Section~\ref{sec:lexecutor}).
Finally, we dive into the domain of mining specifications from execution traces (Section~\ref{sec:spec mining}).

\subsection{Studying Static and Dynamic Call Graphs}
\label{sec:call graphs}

Call graphs are a fundamental building block for various program analyses.
Because of their importance, various algorithms for computing call graphs have been proposed, e.g., for C++~\cite{bacon1996fast}, Java~\cite{tip2000scalable}, Scala~\cite{ali2014constructing}, JavaScript~\cite{Feldthaus2013a}, and Python~\cite{DBLP:conf/icse/SalisSLSM21}.
To better understand the properties of such algorithms, several studies compare the call graphs produced by different analyses with each other, e.g., for C~\cite{murphy1998empirical}, Java~\cite{Reif2019,OnTheRecallICSE2020}, JavaScript~\cite{chakraborty2022automatic}, and WebAssembly~\cite{issta2023-tough-call}.
However, to the best of our knowledge, there is no study that compares dynamic and static call graphs for Python.

The following application of \name{} fills this gap by comparing call graphs generated by PyCG~\cite{DBLP:conf/icse/SalisSLSM21}, a state-of-the-art static call graph analysis for Python, with dynamic call graphs generated by DynaPyt~\cite{fse2022-DynaPyt}.
Because \name{} is ready-to-analyze and already integrated with DynaPyt, obtaining dynamic call graphs of the 50 projects in our benchmark requires relatively little effort.

\subsubsection{Experimental Setup}

A call graph $G = (F, E)$ consists of set $F$ of functions and a set $E$ of edges.
An edge $e \in F \times F$ represents the fact that a caller function invokes a callee function.
If a function is called multiple times by the same caller, the call graph contains only a single between the caller and the callee.
To identify functions, both DynaPyt and PyCG resolve the fully qualified name, including the module name.
For a given Python project, we construct a static call graph $G_{static}$ by applying PyCG to the project's source code.
As mentioned in the documentation of PyCG, we specify the project path and the set of all Python files in the project.
To generate a dynamic call graph $G_{dynamic}$, we use the existing dynamic call graph analysis provided by DynaPyt~\cite{fse2022-DynaPyt}.
To this end, we instrument and then execute the projects in our benchmark, as described in Section~\ref{sec:dynamic analysis}.
We set a time limit of six hours and a memory limit of 60GB for the analysis of a single project. Furthermore, we set a timeout of 30s for each individual test case when run with the DynaPyt analysis.
Moreover, we ignore the results of an analysis on a project in case the analysis itself crashes.

Once we obtain the call graphs $G_{static}$ and $G_{dynamic}$ for the projects in \name{}, we compare them with each other.
For a quantitative comparison, we consider three sets:
(i) the set of call graph edges, (ii) the set of all callers, i.e., nodes with at least one outgoing edge, and (iii) the set of callees, i.e., nodes with at least one incoming edge.
%
For both analyses, we ignore calls made by the underlying unit testing tool and calls made directly inside a test case, because we are interested in the calls made within the analyzed project.
For example, if a test file \code{test\_app.py} contains a test case function called \code{test\_1}, then the call to \code{test\_1} appears neither in $G_{static}$ nor $G_{dynamic}$.
Likewise, any calls made directly inside \code{test\_1} are also ignored.
In contrast, calls made by functions invoked by \code{test\_1} are added to the call graphs.
In addition to the quantitative comparison, we manually inspect a sample of 97 differences between the static and dynamic call graphs.
The goal of this inspection is to understand the differences and to classify them by their root cause.

\subsubsection{Results}

\paragraph{Static call graphs}
PyCG successfully creates a call graph for 39 out of 50 projects.
The missing projects are due to six projects that exceed the six-hour timeout, three projects that exceed the memory limit of 60GB, and two projects where PyCG crashes with an exception. 
Across the 39 successful projects, the static call graphs have a total of 60,565 edges, with an average of 1,552 edges per project.
These edges are between 33,795 unique callers and 10,537 unique callees, where uniqueness is computed at the project level, i.e., if the same function appears in two different projects we count it twice.

\paragraph{Dynamic call graphs}
DynaPyt successfully creates dynamic call graphs for all 50 projects.
On average, executing \name{} while computing the call graph takes 215 minutes per project.
This time is longer than simply executing \name{} without computing call graphs because DynaPyt introduces additional overhead by instrumenting the code.
The dynamic call graphs have a total of 9,575 edges, giving an average of 191 edges per project. These edges are between 3,078 unique callers and 5,384 unique callees. Considering only the 39 projects where PyCG runs successfully, DynaPyt generates 6,306 edges between 2,049 callers and 3,741 callees.

\begin{figure*}[t]
	\captionsetup[subfigure]{aboveskip=-3pt,belowskip=-3pt}
	\begin{subfigure}[b]{0.32\textwidth}
		\centering
		\includegraphics[width=\textwidth]{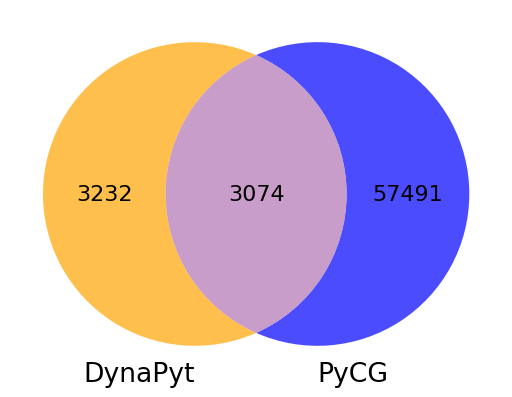}
		\caption{Edges}
		\label{fig:edges}
	\end{subfigure}
	\begin{subfigure}[b]{0.32\textwidth}
		\centering
		\includegraphics[width=\textwidth]{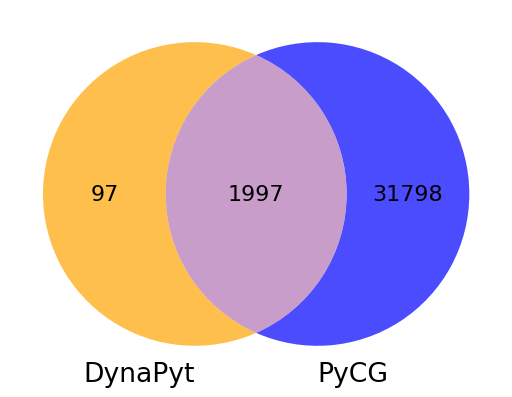}
		\caption{Callers}
		\label{fig:callers}
	\end{subfigure}
	\begin{subfigure}[b]{0.32\textwidth}
		\centering
		\includegraphics[width=\textwidth]{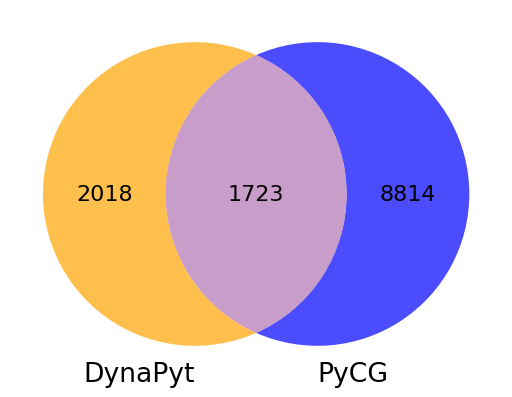}
		\caption{Callees}
		\label{fig:callees}
	\end{subfigure}
	
	\caption{Venn diagrams of edges, callers, and callees in dynamic call graphs (left) and static call graphs (right).}
	
	\label{fig:venncallgraph}
\end{figure*}

\paragraph{Comparison of edges}
Figure~\ref{fig:edges} compares the edges found in the static and dynamic call graphs across the 39 projects that could be successfully analyzed by both the static and dynamic analysis.
The static call graphs have many more edges than their dynamic counterparts, which is expected because the static analysis considers all code of a project, whereas the dynamic analysis focuses on code exercised by the tests in \name{}.
Surprisingly, many edges found by the dynamic analysis are not detected by the static analysis.
Ideally, a static call graph should be sound, i.e., include all call edges that can occur at runtime.
Instead, we find that only 49\% of the edges present within DynaPyt's call graphs are within the set of edges found in PyCG's call graphs. 

\paragraph{Comparison of callers}
To better understand the differences between the static and dynamic call graphs, we compare their sets of callers and callees.
As illustrated in Figure~\ref{fig:callers}, we notice a big difference when comparing the sets of callers derived from PyCG and DynaPyt.
While almost all the callers collected by DynaPyt (95\%) are present in the set of callers identified by PyCG, the static call graph contains many callers missed by the dynamic analysis.
We attribute this difference to the limited code coverage in the exercised tests, which does not exercise all possible behavior of the projects.

\paragraph{Comparison of callees}
When checking the set of callees, an even more important disparity emerges between DynaPyt and PyCG, as shown in Figure~\ref{fig:callees}.
Only 48.5\% of DynaPyt's callees are included in the set of PyCG's callees.
That is, the static call graph is missing many called functions, even though \name{} offers evidence in the form of actual executions that these functions get called.

\paragraph{Root causes of differences}
To understand the reasons that cause the static and dynamic call graphs to differ, we first automatically analyze why edges are missing, and then manually inspect a sample of missing callers and callees.
Our exploration begins with an examination of edges in the dynamic call graph (\dynamic{}) that do not find correspondence in the static call graph (\static{}). Remarkably, 51\% of the edges in \dynamic{} remain unrepresented in \static{}. A detailed analysis of these disparities reveals that for 92\% of such edges, the reason is that the callee is missing in \static{} (while the callers exist). Furthermore, in 8\% of cases, both the callers and callees coexist within \static{}, albeit without an edge connecting them.
We also analyze \static{} edges that do not appear in \dynamic{}. This analysis shows that for 79\% of such edges both the callers and the callees are missing from the dynamic call graph, while for 20\% of these edges only the callees are missing from \dynamic{}.

Next, we manually inspect disparities in callers, more specifically the 5\% of callers appearing in \dynamic{} but not in \static{}.
Our manual inspection shows that the mismatch is caused by differences in the way DynaPyt and PyCG refer to the fully qualified name of some callers. For example, within the Flask project, there is a wrapper around the \code{Requests} library, which makes PyCG resolve the function to \code{FlaskAPI.Requests.something}. Instead, DynaPyt refers directly to the original caller, without the  wrapper, which results in the caller's name being \code{Requests.something}. 

Finally, we analyze the callees captured by \dynamic{} but still missing from \static{}. We identify two primary causes for callees missed by the static analysis:
\begin{itemize}
	\item Our manual inspection shows that PyCG's is missing many calls to methods offered by Python's built-in types. For instance, a call \code{"abc".strip()} is adequately detected by DynaPyt, recording them it as a call to \code{str.strip}, but PyCG fails to capture the call.
	Based on a list of Python's built-in function names, we find that this particular phenomenon accounts for 59\% of all callees missing in the static call graph. In addition, calls in the form of \code{super.method()} are also not included in the call graph of PyCG. 
	\item Another reason is, again, the resolution of modules names, which causes differences in the fully qualified names of functions.
	The Flask example given above illustrates this problem. Differences in function name resolution account for roughly 12\% of the observed mismatches.	
\end{itemize}

\subsubsection{Implications}
To the best of our knowledge, the call graphs generated with \name{} are the first large-scale dataset of dynamic call graphs for Python.
Such a dataset may serve as a ground truth of calls that a sound static analysis should at least detect.
For example, our finding that PyCG is missing method calls made on objects of built-in types, such as strings, calls for considering such calls in future static call graph analyses.
Moreover, the dataset provides a basis for a more detailed empirical study of call graph generation algorithms for Python.
One challenge to be addressed in such a study is to define a uniform way of resolving functions into a fully qualified name, to avoid spurious differences between different call graphs for the same project.
Furthermore, the dynamic call graphs can help evaluate and develop techniques for optimizing and pruning statically generated call graphs~\cite{Utture2022}.
Finally, our observation that dynamic call graphs can be generated quicker than static call graphs for some projects motivates future work on more efficient static call graph algorithms.

\subsection{Gathering Runtime Data for Training a Neural Model}
\label{sec:lexecutor}

Neural models of software are becoming increasingly popular for various applications~\cite{NeuralSoftwareAnalysis}, e.g., code completion, bug detection, and type prediction.
While the majority of approaches focuses on training models on source code and other static artifacts of software, there is an emerging subfield that trains models on data gathered during the execution of programs~\cite{pei2020trex,fse2021,pei2021stateformer,fse2023-LExecutor}.
A major challenge in this line of work is to gather large-scale datasets of runtime data, which is required to obtain effective neural models.
The following demonstrates that \name{} can help address this challenge.

\subsubsection{Experimental Setup}

As a case study, we use LExecutor~\cite{fse2023-LExecutor}, a recent technique that queries a neural model to predict otherwise missing values, and hence, allows for executing arbitrary code snippets.
The neural model underlying LExecutor is trained on a dataset of value-use events.
Each such event consists of a value observed during the execution of a program and the code context in which the value gets used.
The original dataset used to train LExecutor consists of 214,365 value-use events, which were gathered from five projects.

To evaluate the usefulness of \name{} for training LExecutor, we use the benchmark to generate an additional dataset of value-use events.
To this end, we instrument the projects in \name{} with the custom Python instrumentation that is part of LExecutor and then execute the instrumented benchmark, which yields a dataset of value-use events.
We keep a random 5\% of the dataset for validation, and use the remaining 95\% to train a new LExecutor model.
During training, we follow the default setup in the original LExecutor work, i.e. its fine-grained value abstraction and the CodeT5 model, except that we train for five instead of ten epochs due to computational constraints.

\subsubsection{Results}

\begin{figure}
	\includegraphics[width=0.7\linewidth]{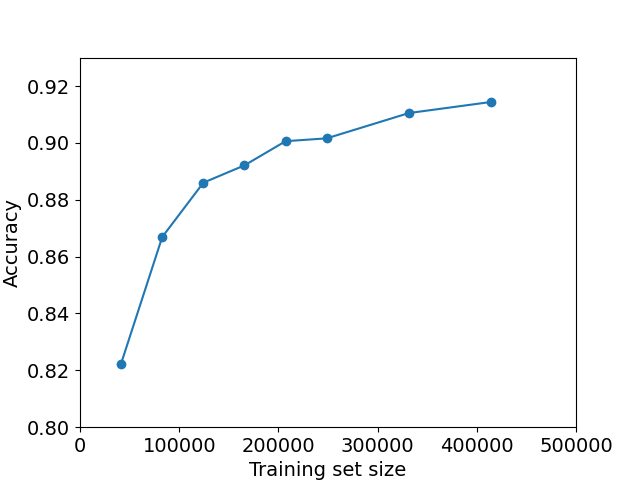}
	\caption{Accuracy of LExecutor~\cite{fse2023-LExecutor} model trained on \name{}-based data.}
	\label{fig:lexecutor_acc}	
\end{figure}

Executing \name{} results in a total of 436,355 value-use events.
That is, thanks to \name{}, the available training data is more than double the size compared to the original LExecutor work.

To assess the impact of using training datasets of different sizes, we train different LExecutor models with increasingly large subsets of the combined training dataset.
Specifically, we train models with subsets of 10\%, 20\%, 30\%, 40\%, 50\%, 60\%, 70\%, 80\%, 90\%, and 100\% of the data, respectively.
We then measure accuracy on the held-out validation data.
Following the original LExecutor work, we report top-1 accuracy, i.e., the percentage of predictions for which the correct value is the most likely prediction.
Figure~\ref{fig:lexecutor_acc} shows the accuracy of the resulting models.
The results illustrate that, perhaps unexpectedly, having more training data leads to a more accurate model, i.e., \name{} provides an easy way to improve the accuracy of the original model.

\subsubsection{Implications}

The continuously increasing interest in neural models of software calls for large-scale datasets of runtime data.
\name{} can help address this need by providing a ready-to-analyze benchmark of executable Python projects.
We envision future work to build upon our benchmark to generate various other kinds of runtime datasets, e.g., for predicting function names from traces of function executions, for predicting static type annotations from types observed at runtime, or for predicting the next value of a variable from its previous values.

\subsection{Mining Specifications from Execution Traces}
\label{sec:spec mining}

Specification mining~\cite{Ammons2002} is a technique that extracts specifications from existing software.
Various techniques for mining specifications have been proposed~\cite{robillard2012automated}, many of which focus on temporal constraints between function calls.
Specification mining plays a role in various applications, including identifying inconsistencies and anomalies, establishing best practices, and gaining a deeper understanding about the common behavior of a given programs.
The following demonstrates that \name{} can be used to mine specifications from execution traces of Python code.

\subsubsection{Experimental Setup}
In our experiment, we leverage call traces generated by DynaPyt to uncover patterns in function calls within the projects in DyPyBench. The process unfolds as follows:

\begin{enumerate}
	\item We customize DynaPyt's call graph analysis to capture traces of function calls, where each trace entry consists of a caller and a callee function.
	
	\item We instrument and execute the projects in \name{} to capture sequences of function calls for each project. Similar to the call graphs analysis, we only instrument and collect traces of the main source code of a project, without instrumenting the test cases themselves.
	
	\item To refine the data, we post-process the call sequences to inline callees into each caller functions, and to remove sequences that are incomplete due to exceptions.
	
	\item We apply an efficient implementation\footnote{\url{https://github.com/chuanconggao/PrefixSpan-py}} of the PrefixSpan algorithm~\cite{han2001prefixspan} to mine frequent patterns from the sequences of functions called within a specific function.
	We give PrefixSpan a list of sequences, each of which corresponds to the functions called within another function, in the order of their execution.
	We apply the mining algorithm to the aggregated set of sequences of all projects in \name{}.
	To reduce the computational cost of the mining algorithm, we consider only sequences of calls that have a length lower or equal to 100 calls, which corresponds to 94\% of all call traces.
	
	\item Finally, we extract the top-100 patterns with a length greater than two. 
\end{enumerate}
Due to time constraints we limit the collection of call traces to 72 hours, which results in traces from 27 projects. While collecting traces using DynaPyt takes a relatively long time, generating the patterns with PrefixSpan is quite fast when invoked with no constraints on the collected patterns. For example, for all the sequences we collect, PrefixSpan is able to generate the top-100 patterns in less than ten seconds. However, this time is highly influenced by the length of sequences, which is the reason why we exclude sequences longer than 100 calls.

\begin{figure}
	\centering
	\includegraphics[width=0.5\linewidth]{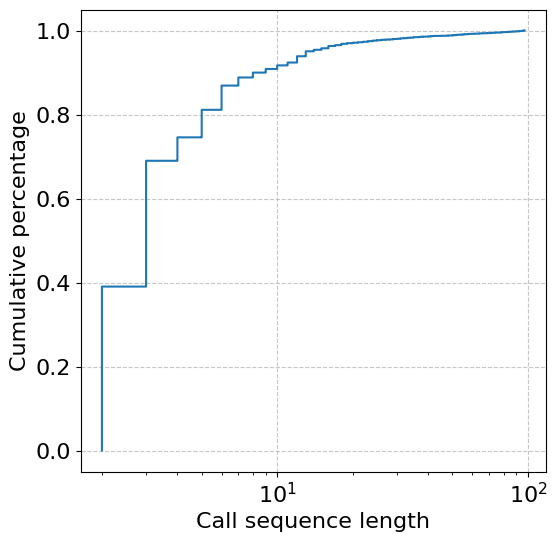}
	\caption{Distribution of extracted calls sequences by length.}
	\label{fig:seqlengthdist}
\end{figure}

\begin{figure}
	\centering
	\includegraphics[width=1.01\linewidth]{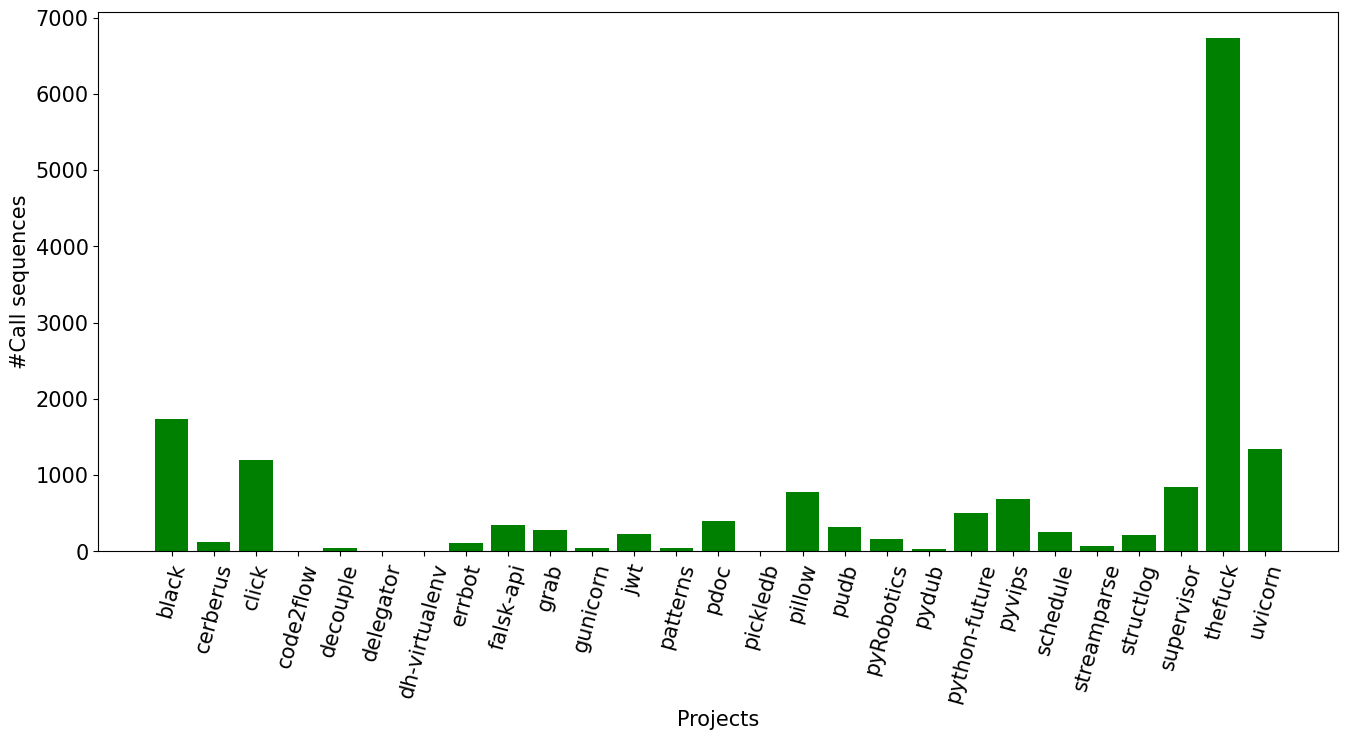}
	\caption{Number of extracted sequences per project.}
	\label{fig:numbersequences}
\end{figure}

\subsubsection{Results}
In total, we extract 16,538 sequences, with an average of 612 sequences per project. The average length is eight, which means that the average function transitively performs eight function invocations.
The CDF in Figure~\ref{fig:seqlengthdist} shows the distribution of the length of the  collected sequences. The plot indicates that most of the sequences (91\%) contain less than ten calls. This implies that the most frequent patterns extracted by PrefixSpan are going to be relatively short patterns.
In addition, the plot in Figure~\ref{fig:numbersequences} illustrates the number of call sequences collected from each project, which ranges between tens of sequences and several thousands. The distribution, however, is far from uniform, which based on our initial results, leads to the patterns extracted from one project dominating the top-100 patterns that we mine from the entire set of projects. To prevent this bias, we set a limit on the number of sequences that we consider from each project during the pattern mining. We define the limit as the mean number of sequences (613) over all projects, plus one standard deviation (1,280), which totals to 1,893 sequences.
Among the top-100 patterns, we randomly pick five patterns and present them in Table~\ref{tab:examples} alongside their frequency and an example of their occurrence in a project in \name{}. The patterns range from sequences of built-in functions and standard library functions (e.g., in rows 1 and 2) to library-specific patterns (e.g., in rows 3 and 4). While the extracted patterns look simple and straightforward, the code reflecting those patterns can be complicated due to multiple branches (e.g., in row 3 and 4) or because they occur in a complex expressions, such as in row 2. Collecting execution traces and mining frequent patterns condenses such behavior into a temporal specification. 

\begin{table}
	\caption{Examples of patterns among top-100 mined patterns.}
	\label{tab:examples}
	\setlength{\tabcolsep}{15pt}
	\begin{tabular}{@{}lrl@{}}
		\toprule
Pattern & \multicolumn{1}{l}{Freq.} & \multicolumn{1}{l}{Code examples from DyPyBench} \\ 
\midrule
\begin{lstlisting}[aboveskip=-.8em,belowskip=-.7em]
builtins.isinstance
builtins.isinstance
\end{lstlisting} & 1,701 & 
\begin{lstlisting}[aboveskip=-.8em,belowskip=-.7em,language=Python]
if isinstance(ty, tuple):
  return Tuple(ty)
if isinstance(ty, ParamType):
  return ty
\end{lstlisting}\\
\hline
\begin{lstlisting}[aboveskip=-.8em,belowskip=-.7em]
bytes.join
builtins.str
str.encode
\end{lstlisting} & 1,008 & 
\begin{lstlisting}[aboveskip=-.8em,belowskip=-.7em,language=Python]
return b"".join([b"HTTP/1.1",
  str(status_code).encode(), 
  b"\s", phrase, b"\r\n"])
\end{lstlisting}\\
\hline
\begin{lstlisting}[aboveskip=-.8em,belowskip=-.7em]
Pattern.match
Match.span
str.isidentifier
\end{lstlisting} & 730 &  \begin{lstlisting}[aboveskip=-.8em,belowskip=-.7em,language=Python]
pseudomatch=pseudoprog.match(line,pos)
if pseudomatch:  # scan for tokens
  start, end=pseudomatch.span(1)
  #code in between
if ...
elif initial.isidentifier():
  # ...
\end{lstlisting}\\
\hline
\begin{lstlisting}[aboveskip=-.8em,belowskip=-.7em]
thefuck...encode_utf8
str.replace
shlex.split
str.replace
thefuck...decode_utf8
\end{lstlisting} & 565 &  \begin{lstlisting}[aboveskip=-.8em,belowskip=-.7em,language=Python]
encoded = self.encode_utf8(command)
try:
  splitted=[s.replace("??", "\\ ") 
  	for s in shlex.split(
  	encoded.replace('\\ ','??'))]
except ValueError:
  splitted = encoded.split(' ')
return self.decode_utf8(splitted)
\end{lstlisting}\\
\hline
\begin{lstlisting}[aboveskip=-.8em,belowskip=-.7em]
logging.getLogger
logger.setLevel
\end{lstlisting} & 205 &  \begin{lstlisting}[aboveskip=-.8em,belowskip=-.7em,language=Python]
logger = logging.getLogger(
	"urllib3.connectionpool")
logger.setLevel(logging.WARNING)
\end{lstlisting}\\
\bottomrule
	\end{tabular}
\end{table}

\subsubsection{Implications}
Despite many research results on specification mining in general~\cite{robillard2012automated}, our work is the first application of specification mining to Python.
\name{} provides a natural foundation for developing and exploring specification mining in Python projects, as it provides a large number of ready-to-run and ready-to-analyze projects.
Beyond specification mining, future work could explore other applications in software analysis that leverage our dataset of call sequences, e.g., to detect anomalies and inconsistencies.

\section{Related Work}

\paragraph{Dynamic analysis and testing of Python}

Python has been the target of several dynamic analyses, e.g.,
to slice programs~\cite{DBLP:conf/compsac/ChenCZXCX14},
to detect type-related bugs~\cite{Xu2016a}, and 
to enforce differential privacy~\cite{DBLP:conf/csfw/AbuahSDN21}.
Scalene~\cite{berger2023triangulating} offers an efficient CPU and memory profiler.
\name{} can serve as a dataset to test and evaluate these and future dynamic analyses.
Pynguin~\cite{Lukasczyk2019} is a unit-level, search-based test generator for Python.
LExecutor~\cite{fse2023-LExecutor} proposes learning-guided execution, which executes code by injecting otherwise missing runtime values.
Both Pynguin and LExecutor share with \name{} the goal of executing code.
\name{} complements these existing efforts by providing a reproducible setup that allows for executing and analyzing a set of real-world projects with little effort.
DynaPyt~\cite{fse2022-DynaPyt} is a general-purpose dynamic analysis framework.
Because it allows for implementing arbitrary dynamic analyses, we integrate DynaPyt into this work and show two applications of it (Sections~\ref{sec:call graphs} and~\ref{sec:spec mining}).

\paragraph{Other work on Python}
Beyond dynamic analysis, the increasing importance of Python has lead to a stream of work on studying Python software and on supporting Python developers.
A study of flaky tests~\cite{DBLP:conf/icst/GruberLK021} relates to this work because they try to automatically execute the test suites of thousands of Python projects.
\name{} differs by focusing on a reusable benchmark, where each project and its setup is carefully checked by a human before being included into the benchmark, whereas the prior work is based on a large-scale, fully automated experiment.
Another difference is the size and characteristics of the projects.
As highlighted in~\cite{DBLP:conf/icst/GruberLK021}, approximately 75\% of their projects have fewer than 10 tests, 60\% exhibit a coverage below 10\%, and over 90\% of the projects consist of less than 10k lines of code.
In contrast, our benchmark contains only 4\% of projects with less than 10 test cases, a minimum coverage of 48\% (average: 82\%), and an average code base size of 14k lines per project.

The lack of static type annotations by default has lead to several techniques for inferring and predicting type annotations~\cite{Xu2016,fse2020,Allamanis2020,Yan2023}, to work on comparing gradual type systems for Python~\cite{Rak-amnouykit2020}, and to work on studying type annotations in the Python ecosystem~\cite{fse2022_type_study}.
\name{} could be used to create a ground truth of types observed at runtime, which can serve as training data for learning-based type predictors or be used during their evaluation.
Another study investigates the performance benefits of using Python idioms~\cite{Zhang2023}.
Future work on studying the performance of Python code can benefit from \name{} as a ready-to-execute collection of real-world code.

\paragraph{Dynamic analysis of other languages}

Beyond Python, there are various dynamic analyses, e.g., to detect concurrency bugs~\cite{Savage1997,OCallahan2003,Flanagan2004}, type-related problems~\cite{An2011,icse2015}, and other common bug patterns~\cite{issta2015}.
Other analyses find optimization opportunities~\cite{Xu2009,oopsla2015}, or infer API usage protocols~\cite{Yang2006,ase2009} and input grammars~\cite{Hoeschele2016}.
Security-oriented analyses include taint analysis~\cite{Clause2007} and dynamic detectors of similar functions~\cite{Egele2014}.
%
To reduce the effort of building a dynamic analysis, dynamic analysis frameworks have been proposed for most popular languages, e.g., 
DynamoRIO~\cite{bruening2003infrastructure}, Pin~\cite{luk2005pin}, and Valgrind~\cite{Nethercote2007}, which all target x86 binaries,
Jalangi~\cite{Sen2013} for JavaScript,
Wasabi~\cite{asplos2019} for WebAssembly,
DiSL~\cite{DBLP:conf/aosd/MarekVZABQ12} for Java, and RoadRunner~\cite{Flanagan2010}, which specifically targets concurrency-related dynamic analyses.
As \name{} targets Python, it cannot directly support the above analyses, but the breadth and depth of prior work underlines the importance of dynamic analysis, and hence, benchmarks like ours.

\paragraph{Benchmarks of executable code}

Our work has been partially inspired by other benchmarks of executable code.
Similar to \name{}, several benchmarks offer a set of real-world programs with inputs to execute their code, e.g., DaCapo~\cite{Blackburn2006} for Java, SPEC CPU~\cite{Henning2006} for C/C++, and Da Capo con Scala~\cite{sewe2011capo}.
Other executable benchmarks are aimed at specific tools and techniques, e.g., LAVA~\cite{Dolan-Gavitt2016} and Magma~\cite{hazimeh2020magma}, which are aimed at evaluating fuzzers,
SecBenchJS~\cite{icse2023-SecBenchJS}, which is aimed at evaluating techniques for detecting or mitigating vulnerabilities in JavaScript, and
Parsec, which offers multi-threaded C/C++ programs ~\cite{bienia11benchmarking}.
Importantly, none of the above benchmarks is for Python, which is the gap our work tries to fill.

\paragraph{Other benchmarks}

Many other benchmark sets have proven valuable, of which we can list only a few here.
To evaluate and compare bug-related techniques, there are Defects4J~\cite{Just2014}, Bugs.jar~\cite{saha2018bugs}, and BugSwarm~\cite{DBLP:journals/corr/abs-1903-06725} for Java, BugsJS for JavaScript~\cite{gyimesi2019bugsjs}, and BugsInPy~\cite{widyasari2020bugsinpy} for Python.
More specialized benchmarks include a set of JavaScript performance bugs~\cite{icse2016-perf} and a collection of concurrency bugs~\cite{yuan2021gobench}.
To evaluate deep learning models of code, benchmarks such as CodeXGLUE~\cite{ lu2021codexglue} and HumanEval~\cite{chen2021evaluating}, are widely used.
We envision \name{} to help support future work on dynamic analysis for Python in a way similar to the above benchmarks.

\section{Conclusion}

This paper addresses a significant gap in the Python programming ecosystem by introducing \name{}, a dynamic benchmark suite of executable Python projects.
The benchmark offers a unique combination of features by being large-scale, diverse, ready-to-run, and ready-to-analyze.
With 50 projects, 681k lines of code (of which 558k are executed), and 29,511 test cases, \name{} offers a rich set of executions to analyze and study.
We illustrate the practicality and utility of \name{} in three usage scenarios: comparing static and dynamic call graphs; creating training data for learning-guided execution; and mining specifications from execution traces.
We envision our work to provide a foundation for many other dynamic analyses and for studying the runtime behavior of Python software.

\section*{Data availability}
Our DyPyBench image is available on Zenodo: \url{https://zenodo.org/doi/10.5281/zenodo.10683759}.
Alternatively, the image can also be pulled from DockerHub: \url{https://hub.docker.com/r/islemdockerdev/dypybench}.
Project page for updates and issues: \url{https://github.com/sola-st/DyPyBench}

\section*{Acknowledgments}
This work was supported by the European Research Council (ERC, grant agreement 851895), and by the German Research Foundation within the ConcSys, DeMoCo, and QPTest projects.

\bibliographystyle{ACM-Reference-Format}
\bibliography{paper-dypybench,referencesMP,referencesMore}
\end{document}